\documentclass[12pt]{iopart}

\usepackage[pdftex]{graphicx}
\usepackage[pdftex]{color}

\usepackage{iopams}

\expandafter\let\csname equation*\endcsname\relax 
\expandafter\let\csname endequation*\endcsname\relax

\usepackage{amsmath}
\usepackage{mathrsfs}
\usepackage{citesort}

\definecolor{darkgreen}{rgb}{0.0, 0.5, 0.0}
\definecolor{purple}{rgb}{0.50, 0.0, 0.50}

\def\dps{\displaystyle}
\def\To{\Rightarrow}

\def\sn{\mathrm{sn}}
\def\mrm{\mathrm}

\def\del{\partial}

\def\vep{\varepsilon}

\newcommand{\argmax}{\mathop{\rm arg~max}\limits}

\def\pb{\overline{p}}

\def\FF{\vec{F}}

\def\HH{\vec{H}}
\def\vg{\vec{g}}
\def\ff{\vec{\mathcal{R}}}
\def\uu{\vec{u}}
\def\vv{\vec{v}}

\def\GGG{\mathcal{G}}
\def\MMM{\mathcal{M}}

\def\HHH{\vec{\mathcal{H}}}
\def\UUU{\vec{\mathcal{U}}}

\def\rmV{\mathrm{V}}
\def\rmE{\mathrm{E}}
\def\rmI{\mathrm{I}}

\def\pcone{{p_\mrm{c1}}}
\def\pctwo{{p_\mrm{c2}}}

\def\Ncp{{N_\mrm{cp}}}
\def\Nrt{{N_\mrm{rt}}}
\def\Ncnc{{N_\mrm{cnc}}}
\def\Nsgl{{N_\mrm{uni}}}
\def\ycp{{y_\mrm{cp}}}

\def\be{ \begin{eqnarray} }
	\def\ee{ \end{eqnarray} }

\newcommand{\beq}{\begin{equation}}
	\newcommand{\eeq}{\end{equation}}

\begin{document}

\title
[Renormalization-group theory of the abnormal singularities ...]
{Renormalization-group theory of the abnormal singularities 
	at the critical-order transition in bond percolation on pointed hierarchical graphs}

\author{Tomoaki Nogawa}

\address{$^1$ Faculty of Medicine, Toho University, 5-21-16, Omori-Nishi, Ota-ku, Tokyo, 143-8540, Japan}
\ead{nogawa@med.toho-u.ac.jp}
\vspace{10pt}
\begin{indented}
\item[] August 2017
\end{indented}

\begin{abstract}
We study the singularity of the order parameter at the transition 
between a critical phase and an ordered phase 
of bond percolation on pointed hierarchical graphs (PHG). 
In PHGs with shortcuts, 
the renormalization group (RG) equation explicitly depends on the bare parameter, 
which causes the phase transition that corresponds to the bifurcation of the RG fixed point. 
We derive the general relation between the type of this bifurcation 
and the type of the singularity of the order parameter. 
In the case of a saddle node bifurcation, 
the singularity is power-law or essential type 
depending on the fundamental local structure of the graph. 
In the case of pitchfork and transcritical bifurcations, 
the singularity is essential and power-law types, respectively. 
These are replaced by power-law and discontinuous types, respectively, 
in the absence of the first-order perturbation to the largest eigenvalue of the combining matrix, 
which gives the growth rate of the cluster size.
We also show that the first-order perturbation vanishes 
if the backbone of the PHG is simply connected via nesting subunits 
and all the roots of the PHG are almost surely connected in the ordered phase. 
\end{abstract}

%
%
%
%
%

\section{Introduction}

Recently, cooperative phenomena on non-Euclidean graphs 
have been extensively studied in the context of complex networks \cite{Dorogovtsev08}. 
Such systems sometimes show behaviors quite different from those of the Euclidean systems 
due to small-worldness, i.e., infinite dimensionality, 
and hierarchical structures coming from the growth mechanism \cite{Ravasz-Barabasi2003}. 
One of the interesting topics on such systems is persistent criticality, i.e., 
the systems show the power-law properties, that are observed at the critical point 
of the second-order transitions, in a finite-volume region of the parameter space. 
We call such a region a critical phase \cite{Hasegawa-Nogawa-Nemoto14}. 
This phase is similar to the so-called Berezinskii-Kosterlitz-Thouless (BKT) phase 
\cite{Berezinskii72,Kosterlitz73,Kosterlitz74}, 
which is observed in the system at the lower critical dimension 
where the long-range coherence of continuous order parameter is marginally unstable. 
But the studies on hierarchical small-world networks (HSWNs)
has been revealed that the origin of persistent criticality 
is quite different between the two in the renormalization group (RG) aspect. 
In the RG theory, the model parameter (vector in general) assigned on the edges or vertices, 
such as the open-bond probability of percolation, are transformed through coarse-graining. 
By repeating the transformation, 
the parameter converges to one of the fixed points (FPs)
depending on the initial condition, i.e., the bare parameter. 
The set of the bare parameters that converge to the same stable FP form a phase, 
and the parameters at a phase boundary converge to a saddle point. 
The BKT phase is represented by the fixed {\it line}, i.e., the array of FPs. 
While the parameter is renormalized homogeneously in space in the ordinary RG theory, 
it is renormalized inhomogeneously in HSWNs; 
the parameters on the backbone edges are transformed 
but those on the shortcut edges are not \cite{Boettcher-Brunson15}. 
In the former case, the RG equation itself does not depend on the bare parameters, 
which only play the role of the initial condition. 
In the latter case, the RG equation explicitly depends on the bare parameter. 
The critical phase of HSWNs is represented by a single FP which moves as the bare parameter changes.
Furthermore, such a FP possibly undergoes a bifurcation by changing the bare parameter, 
which interlocks a phase transition.

The singularity of the order parameter at critical-order transition is also an interesting issue. 
Several abnormal singularities are observed. 
(In the case of percolation problem, an ordered phase corresponds to 
a percolating phase where an extensively large cluster exists. )
First type is the essential singularity. 
This is observed in the so-called inverted BKT transition \cite{Hinczewski06}, 
which is similar to the dual version of the BKT transition 
in the solid-on-solid model \cite{Chui-Weeks1976} 
and the $p$-state clock model with $p \ge 5$ \cite{Jose1977}. 
This type has been reported in relatively many systems: percolation 
\cite{Krapivsky04,Bollobas05,Riordan05,Berker09,Hasegawa-Sato2010,Boettcher09,Hasegawa-Nemoto10}
and spin systems 
\cite{Bauer05,Hinczewski06,Boettcher11,Nogawa-Hasegawa12,Nogawa-Hasegawa12b}. 
Second type is the power-law singularity that is governed not by a saddle FP 
but by a stable FP unlike the ordinary second-order transitions \cite{Nogawa-Hasegawa12b}. 
Third type is an abrupt singularity, where the order parameter changes discontinuously \cite{Boettcher12}. 
Similar discontinuous transition is observed in the numerical simulation 
of the hyperbolic lattice \cite{Nogawa-Hasegawa09}, 
which has a dual relation to the so-called infinite-order transition known in the Cayley trees 
\cite{Eggarter74,Muller-Hartmann74,Ostilli12,Nogawa-Hasegawa-Nemoto16}. 
All of these types of singularity are observed in the single system, 
where the two types of the graph-growth rules are randomly mixed, 
by tuning the mixing ratio \cite{Nogawa-Hasegawa14}. 
Previous studies also imply that the type of the singularity of the order parameter 
corresponds to the type of the bifurcation of the RG FP; 
essential, power-law and abrupt singularities are related to 
the saddle-node, pitchfork and transcritical bifurcations, respectively. 
The reason for this, however, has not been revealed yet. 
In this paper, we provide a general scheme to determine 
the singularity of the order parameter for a given RG equation.

This paper is organized as follows. 
In the next section, we introduce the mathematical settings for the analysis. 
In Sec.3, we argue the general relation between the RG behavior and the order parameter. 
In Sec.4, we derive the singularities of the order parameter 
for the individual types of the RG equation. 
Finally, the results are summarized.


\section{Preliminaries}

\subsection{pointed hierarchical graph (PHG)}

\begin{figure}[t]
	\begin{center}
		\includegraphics[trim=0 20 0 20, scale=0.5,clip]{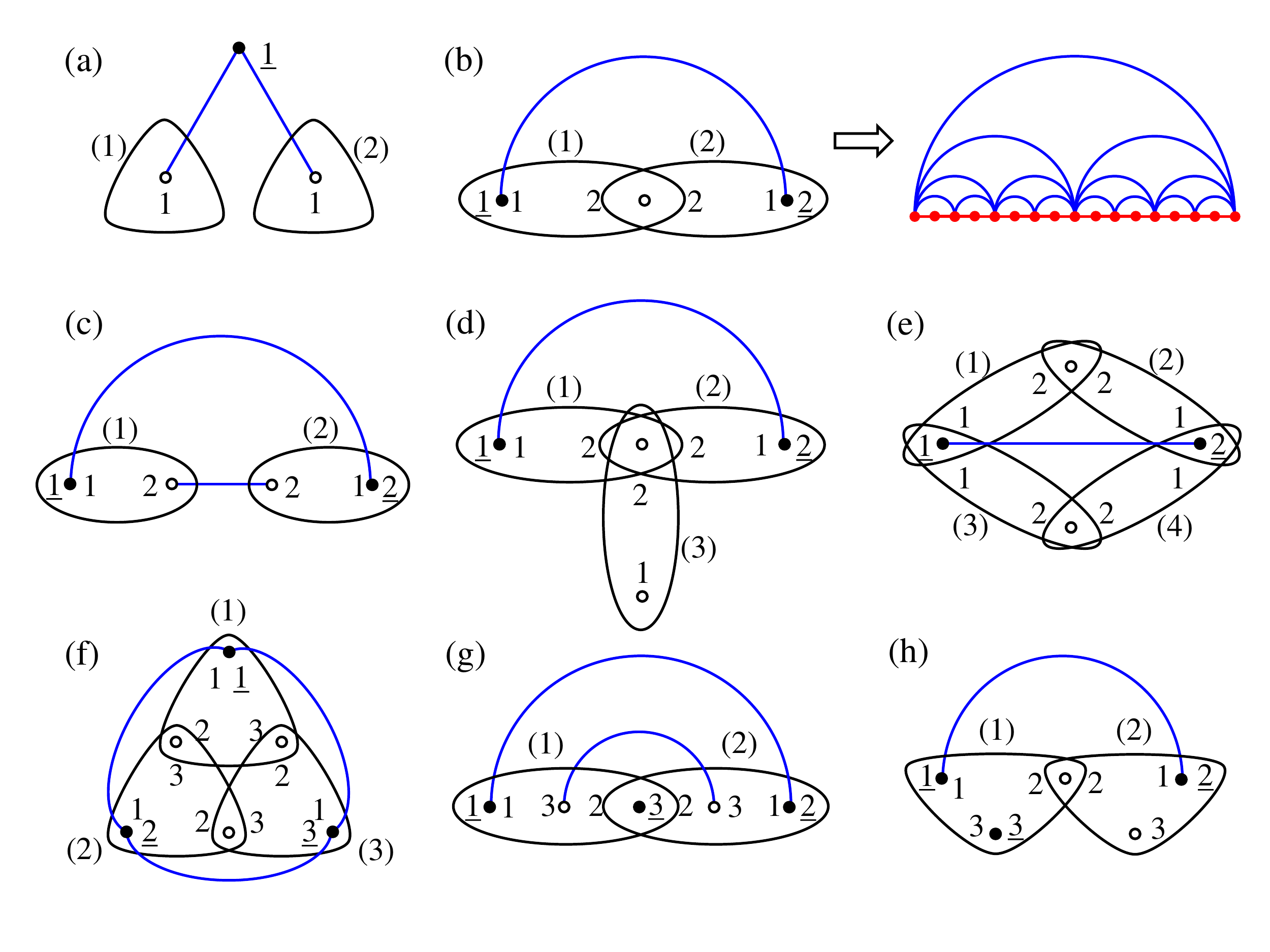}

		\caption{\label{fig:graph}
			Schematic diagrams of the construction of a PHG $\GGG_{n+1}$ from $\GGG_n^{(\cdot)}$'s. 
			In the case of (b), we show $\GGG_3$ together for the case that $\GGG_0$ is connected two vertices. 
			The closed curves denote $\GGG_n^{(\cdot)}$'s 
			and the small circles inside them denote $R_n^{(\cdot)}$'s, 
			some of which are shared by two copies due to identification. 
			Particularly, the filled circles denote $R_{n+1}$. 
			The numbers with round brackets are the indices of $\GGG_n^{(\cdot)}$'s. 
			The numbers without and with underline are the indices of $R_n^{(\cdot)}$'s 
			and $R_{n+1}$, respectively. 
			The graph operations and the assignment of the new roots for each PHG are listed below, where 
			$\rmV(v_1)$ means adding a vertex $v_1$, 
			$\rmE(v_1, v_2)$ means adding an edge between vertices $v_1$ and $v_2$, and  
			$\rmI(v_1, v_2)$ means identifying vertices $v_1$ and $v_2$. 
			\\
			(a) $\rmV(v_1),\ \rmE(v_1,r_{n1}^{(1)}),\ \rmE(v_1, r_{n1}^{(2)}),\ 
			R_{n+1}=\{v_1 \}$\\
			(b) $\rmI(r_{n2}^{(1)},r_{n2}^{(2)}),\ \rmE(r_{n1}^{(1)}, r_{n1}^{(2)}),\ 
			R_{n+1}=\{r_{n1}^{(1)}, r_{n1}^{(2)} \}$ \\
			(c) $\rmE(r_{n2}^{(1)},r_{n2}^{(2)}),\ \rmE(r_{n1}^{(1)}, r_{n1}^{(2)}),\ 
			R_{n+1}=\{r_{n1}^{(1)},r_{n1}^{(2)} \}$ \\
			(d) $\rmI(r_{n2}^{(1)},r_{n2}^{(2)}),\ \rmI(r_{n2}^{(1)}, r_{n2}^{(3)}),\ \rmE(r_{n1}^{(1)}, r_{n1}^{(2)}),\ 
			R_{n+1}=\{r_{n1}^{(1)},r_{n1}^{(2)} \}$ \\
			(e) $\rmI(r_{n1}^{(1)}, r_{n1}^{(3)}),\ \rmI(r_{n1}^{(2)}, r_{n1}^{(4)}),\ 
			\rmI(r_{n2}^{(1)}, r_{n2}^{(2)}),\ \rmI(r_{n2}^{(3)}, r_{n2}^{(4)}),\ \rmE(r_{n1}^{(1)}, r_{n1}^{(2)}),\ 
			R_{n+1}=\{r_{n1}^{(1)},r_{n1}^{(2)} \}$ \\
			(f) $\rmI(r_{n2}^{(1)},r_{n3}^{(2)}),\ \rmI(r_{n2}^{(2)},r_{n3}^{(3)}),\ \rmI(r_{n2}^{(3)},r_{n3}^{(1)}),\ 
			\rmE(r_{n1}^{(1)},r_{n1}^{(2)}), \rmE(r_{n1}^{(2)},r_{n1}^{(3)}), \rmE(r_{n1}^{(3)},r_{n1}^{(1)})$, \\
			\quad \quad $R_{n+1}=\{r_{n1}^{(1)}, r_{n1}^{(2)}, r_{n1}^{(3)} \}$ \\
			(g) $\rmI(r_{n2}^{(1)}, r_{n2}^{(2)}),\ \rmE(r_{n1}^{(1)}, r_{n1}^{(2)}),\ \rmE(r_{n3}^{(1)}, r_{n3}^{(2)}),\ 
			R_{n+1}=\{r_{n1}^{(1)}, r_{n1}^{(2)}, r_{n2}^{(1)} \}$ \\
			(h) $\rmI(r_{n2}^{(1)},r_{n2}^{(2)}),\ \rmE(r_{n1}^{(1)}, r_{n1}^{(2)}),\ 
			R_{n+1}=\{r_{n1}^{(1)}, r_{n1}^{(2)}, r_{n3}^{(1)} \}$  \\
		}
	\end{center}
\end{figure}

Let us define a PHG
\footnote{
	A pointed graph is a graph together with a distinguished vertex.
	Here we consider the generalized case that the pointed vertex can be multiple. 
	The class of the PHG includes HSWNs as well as large-world graphs. 
}
, which is an increasing sequence of graphs as 
\be
\GGG_n = \{ V_n, E_n \}, \quad n \in \{0, 1, 2, \cdots \}, \quad  \GGG_n \subset \GGG_{n+1}, 
\ee
where $V_n$ and $E_n$ are the sets of vertices and edges, respectively. 
The pointed members of $V_n$ with serial numbers 
\be
R_n := \{ r_{n1}, \cdots, r_{n \Nrt} \} \subseteq V_n, 
\quad \Nrt \in \{1, 2, \cdots \} 
\ee
are called the roots of $\GGG_n$. 
A PHG is recursively constructed; $\GGG_{n+1}$ is made from $\GGG_n$ as follows. 
\begin{enumerate}
	\item Prepare $\Ncp \in \{2,3,\cdots \}$ copies of $\GGG_n$: $\{\GGG_n^{(1)}, \cdots, \GGG_n^{(\Ncp)}\}$. 
	
	\item Perform a finite number of graph-operations on the roots of the copies, 
	$R_n^{(1)} \cup \cdots \cup R_n^{(\Ncp)}$, 
	such as identifying two vertices and adding vertices or edges. 
	
	\item Choose $\Nrt$ vertices to be $R_{n+1}$ from $R_n^{(\cdot)}$'s and the added vertices. 
\end{enumerate}
Note that a PHG is constructed deterministically 
and its local structure depends on the initial graph $\GGG_0$. 
Some examples of the PHG are shown in figure~\ref{fig:graph}. 
Let us call the edges originally included in $\GGG_0$ {\it backbone} edges 
and call the other edges added in the graph-operations {\it shortcut} edges. 

Next, we define some properties of PHGs with $\Nrt \ge 2$. 
Each $\GGG_n$ (accurately $\GGG_n \setminus R_n$) plays a role of a hyperedge in $\GGG_{n+1}$. 
\begin{itemize}
	\item
	We say that a PHG $\GGG_n$ is {\it open} 
	when there exists at least one path via $e \in E_n$ between all the pairs of the roots in $R_n$. 
	We say that a PHG $\GGG_n$ is {\it closed} when there exists no path between any pairs of the roots in $R_n$. 
	Note that `closed' and `not open' are not equivalent unless $\Nrt=2$.
	
	\item
	We say that a PHG $\GGG_n$ is {\it backbone-connected} when it satisfies the following. 
	If all copies $\GGG_{n-1}^{(\cdot)} \subset \GGG_n$ are open and all shortcut edges are removed, $\GGG_n$ is open. 
	
	\item
	We say that a backbone-connected PHG $\GGG_n$ is {\it simply-backbone-connected (SBC)} when it satisfies the following. 
	If one arbitrary copy $\GGG_{n-1}^{(\cdot)} \subset \GGG_n$ is closed and all shortcut edges are removed, $\GGG_n$ is not open.
\end{itemize}
In table~\ref{tab:graph}, the properties of the PHGs in figure~\ref{fig:graph} are listed. 
The PHG (c) is not backbone-connected 
because $\GGG_{n}^{(\cdot)}$'s are connected not by identification but by adding an edge. 
The PHG (d) and (e) are not SBC 
because they have a dangling dead-end and a redundant path, respectively. 
For $\Nrt=2$, the backbone structure of a SBC PHG is unique, 
that is, a one-dimensional chain whose two end points are the new roots. 

\begin{table}[t]
	\begin{center}
		\begin{tabular}{l|cc|ccc|ccl}
			\hline
			name & $\Nrt$ & $\Ncp$ & BC & SBC & TRC & BF & $\mu$ & singularity \\
			\hline
			\begin{tabular}{l} (a) binary tree  \end{tabular}
				& 1 & 2 & -- & -- & -- & -- & -- & none \\
			\begin{tabular}{l} (b) Farey graph \cite{Zhang-Comellas2011}  \end{tabular}
				& 2 & 2 & T & T & T & TC & 2 & abrupt \cite{Boettcher12}\\
			\begin{tabular}{l} (c) {\it nameless}  \end{tabular}
				& 2 & 2 & F & -- & -- & -- & -- & -- \\
			\begin{tabular}{l} (d) {\it nameless}  \end{tabular}
				& 2 & 3 & T & F & T & TC & 1 & power-law \\ \hline
			\begin{tabular}{l} (e) decorated \\ \quad \ (2,2)-flower \cite{Hinczewski06} \end{tabular}
				& 2 & 4 & T & F & T & SN & -- & essential \cite{Rozenfeld07,Hasegawa-Sato2010} \\
			\begin{tabular}{l} (f) Sierpinski Gasket \\ \quad \ with shortcuts  \end{tabular}
				& 3 & 3 & T & T & T & SN & -- & none \\
			\begin{tabular}{l} (g) Hanoi network \\ \quad \ (planar) \cite{Boettcher09}  \end{tabular}
				& 3 & 2 & T & T & T & TC & 2 & abrupt \\
			\begin{tabular}{l} (h) {\it nameless}  \end{tabular}
				& 3 & 2 & T & T & F & TC & 2 & abrupt \\
			\hline
		\end{tabular}
	\end{center}
	\caption{\label{tab:graph}
		Properties of PHGs in figure~\ref{fig:graph} and bond percolation on them. 
		The columns `BC' and `SBC' denotes whether it is true(T) or false(F) that 
		each PHG is backbone-connected and simply-backbone-connected, respectively.
		The column `TRC' denotes whether the PHG has a tight-root-connection or not.  
		The column `BF' denotes the type of the bifurcation of the RG fixed point at $p=\pctwo$; 
		TC, SN and `--' denote transcritical, saddle-node and no bifurcation ($\pctwo=1$), respectively. 
	}
\end{table}

\subsection{bond percolation on PHG}

Let us consider bond percolation on PHGs; 
each edge is open or closed probabilistically and independently. 
In this paper, we suppose that 
all shortcut and backbone edges are open with a unique probability $p$ 
and closed with the probability $\pb := 1-p$. 
Although most of the quantities appearing hereafter are the functions of $p$, 
we do not denote it explicitly.

Let $C_{nv} \subseteq V_n$ be the connected component of the open edges 
for a given open-edge realization that includes $v\in V_n$. 
Then we define the order parameter $m$ and the fractal exponent $\psi$ as 
\be 
m := \lim_{n \to \infty} \frac{\sigma_n}{|V_n|}, 
\quad 
\psi := \lim_{n \to \infty} \frac{\ln \sigma_n}{\ln |V_n|}, 
\quad 
\nonumber \\
\sigma_n := E(|C_{no}|), 
\quad 
o := \mathop{\argmax}_{r \in R_n} E(|C_{nr}|), 
\ee
where $E(\cdots)$ denotes the expectation value 
and $|\cdots|$ denotes the number of the members of a set. 
Then, $|C_{nv}|$ means the size of a cluster. 
The number of the vertices increases as $|V_n| \propto \Ncp^n$ for $n \to \infty$. 
We call $o$ the origin of $\GGG_n$ and 
suppose that $o=r_{n1}$ for $n \gg 1$. 
By using these quantities, two critical probabilities are defined as 
\be
\pcone &:=& \sup \{ p | \psi = 0 \}, 
\nonumber \\
\pctwo &:=& \inf \{ p | \psi =1 \} = \inf \{ p | m>0 \}. 
\ee
These coincide in percolation on Euclidean lattices, 
but they often do not in percolation on hierarchical small-world graphs. 
We call the regions $[0,\pcone)$, $(\pctwo,1]$ and $(\pcone, \pctwo)$ 
a disordered phase, an ordered phase and a critical phase, respectively. 

\subsection{generating functions}
\label{sec:gen_fnc}

Let $\Omega_{\Nrt} = \{ \omega_1, \cdots, \omega_{\Ncnc} \}$ be 
the set of the possible connectivities among the $\Nrt$ roots 
and denote ``$R_n \sim \omega$'' when the connectivity of $R_n$ is given by $\omega$. 
For example, we express ``$R_n \sim [1,2,1]$'' when $r_{n1}$ and $r_{n3}$ 
are in the same cluster (1) and $r_{n2}$ is in another cluster (2). 
We have $\Omega_2$ = \{[1,1], [1,2]\} and 
$\Omega_3$ = \{[1,1,1], [1,2,2], [1,2,1], [1,1,2], [1,2,3]\}, and so on.
Here $\Ncnc$ is expressed by the Stirling numbers of the second kind $S(k,j)$ as 
\be
\Ncnc := |\Omega_{\Nrt}| 
= \sum_{j=1}^{\Nrt} S(\Nrt,j), \quad 
S(k,j) := \frac{1}{j!} \sum_{i=1}^j (-1)^{j-i} \left( \begin{array}{c} j \\ i \end{array} \right) i^k. 
\ee
Let $a_\omega$ be the number of the distinct clusters that include a root in $\omega \in \Omega_\Nrt$, 
and we define $A_{\Nrt} = \{a_{\omega_1}, \cdots, a_{\omega_{\Ncnc}} \}$, 
e.g., $A_2$ = \{1, 2\} and $A_3$ = \{1, 2, 2, 2, 3\}.

The generating function corresponding to $R_n \sim \omega$ is defined as 
\be
G_{n\omega}(x_1, \cdots, x_{a_\omega}) := 
\sum_{s_1=1}^\infty \cdots \sum_{s_{a_\omega}=1}^\infty x_1^{s_1} \cdots x_{a_\omega}^{s_{a_\omega}}
\nonumber \\ \hspace{2cm} \times
P(R_n \sim \omega \wedge |C_{n1}|=s_1 \wedge \cdots \wedge |C_{na_\omega}|=s_{a_\omega}) 
. 
\ee
Hereafter $P(A)$ denotes the probability of a proposition $A$ is true 
and `$\wedge$' denotes logical conjunction. 
Each argument $x_a$ is respectively related to the cluster that includes at least one root.  
By using the derivatives of these generating functions, we have 
\be
\sigma_n = \sum_{\omega \in \Omega_\Nrt} \left. \frac{d}{dx_1}  G_{n \omega}(x_1, 1, \cdots, 1) \right|_{x_1=1}. 
\ee

Each generating function yields $2^{a_\omega}-1$ univariable functions 
by substituting either $x$ or 1 into each variable, 
e.g., $G(x_1,x_2,x_3)$ yields 
$G(x,x,x)$, $G(1,x,x)$, $G(x,1,x)$, $G(x,x,1)$, $G(x,1,1)$, $G(1,x,1)$ and $G(1,1,x)$. 
Let us define the vector whose components are given by the all possible univariable functions as 
\be
\HH_n(x) = {}^t(H_{n1}(x), \cdots, H_{n\Nsgl}(x) ) , \quad
\Nsgl := \sum_{\omega \in \Omega_\Nrt} [ 2^{a_\omega} - 1 ], 
\ee
where $t$ on the left shoulder denotes transposition of a vector. 
We set the first component of $\HH_n(x)$ to correspond to the open $\GGG_n$, namely, 
$H_{n1}(x)=G_{n \omega_1}(x)$. 
(This is the unique generating function that is originally univariable.) 
Note that $\HH_n(x)$ does not include the constants such as $G(1,1,1)$. 
Instead, they form another vector: 
\be
\vg_n = {}^t(g_{n1},\cdots, g_{n \Ncnc}) = {}^t(G_{n \omega_1}(1), G_{n \omega_2}(1, 1), \cdots, G_{n \omega_\Ncnc}(1, \cdots, 1) ). 
\ee
Note that $g_{n i} = P(R_n \sim \omega_i)$ and then $\sum_{i=1}^{\Ncnc} g_{ni} = 1$.

\subsection{combining matrix}
\label{sec:comb}

For general PHGs, we can express $\{ G_{n+1, \omega}(\cdot)\}$ by $\{ G_{n \omega}(\cdot)\}$
owing to that the connectivities inside the individual copies are independent. 
Furthermore, we can derive the recursion equations for $\vg_n$ and $\HH_n$ as 
\footnote{
	Rigorously, $\FF(\cdot,\cdot)$ would explicitly depend on $x$ not only via $\HH_n(x)$. 
	For example, addition of a vertex yields a factor $x$. 
	We omit such a factor because it only adds an constant vector to the r.h.s. of equation~\eqref{eq:dHdx} 
	and does not change the combining matrix $M_n$.
}
\be
\vg_{n+1} = \ff(\vg_n), \quad
\HH_{n+1}(x) = \FF \left( \vg_n, \HH_{n}(x) \right).
\ee
The former equation is closed in $\{\vg_n\}$. 
We call it a RG equation. 
Note that $\ff(\cdot)$ and $\FF(\cdot,\cdot)$ explicitly depend on $p$ 
if shortcut edges exist.

The derivative of $\HH_n(x)$, which enables us to evaluate $\sigma_n$, 
obeys to a linear recursion equation: 
\be
\left. \frac{d \HH_{n+1}(x)}{dx} \right|_{x=1}
= M_n \left. \frac{ d \HH_n(x)}{dx}  \right|_{x=1}, \quad 
M_{n,ij} := \frac{\del F_i\left(\vg_n, \HH_n(1) \right)}{\del H_{nj}(1)}.  
\label{eq:dHdx}
\ee
We call $M_n$ a {\it combining matrix}, which is generally asymmetric. 
The $k$th largest eigenvalue of $M_n$ is denoted by $\lambda_n^{(k)}$, 
and the conjugating left and right eigenvectors are denoted by 
$\uu_n^{(k)}$ and $\vec{v}_n^{(k)}$, respectively. 
The largest eigenvalue yields the fractal exponent as 
\be
\sigma_n \propto (\lambda_\infty^{(1)})^n 
\quad \mrm{for} \quad n \gg 1 
\quad \To \quad 
\psi = \log_{N_\mrm{cp}} \lambda_\infty^{(1)}. 
\ee

\subsection{example: Farey graph}
\label{sec:farey1}

Here we actually calculate the quantities defined above 
in the case of the Farey graph, the PHG (b) in figure~\ref{fig:graph}, 
where $\Ncp=2$, $\Nrt=2$, $\Ncnc=2$ and $\Nsgl=4$. 
This calculation is essentially same with the results in Ref.~\cite{Boettcher12}. 
The recursion equations for the generating functions 
\be
G_{n1}(x_1) &:=& \sum_{s_1=1}^\infty P(R_n \sim [1,1] \wedge |C_{n1}|=s_1) {x_1}^{s_1} , 
\nonumber \\
\hspace{-6mm} 
G_{n2}(x_1,{x_2}) &:=& \sum_{s_1=1}^\infty \sum_{{s_2}=1}^\infty P(R_n \sim [1,2] \wedge |C_{n1}|=s_1 \wedge |C_{n2}|={s_2}) {x_1}^{s_1} {x_2}^{s_2}, 
\ee
are given by 
\be
\hspace{-6mm} 
G_{n+1,1}(x_1) &=& p \left[  
G_{n1}(x_1) G_{n1}(x_1) + G_{n2}(x_1,x_1) G_{n1}(x_1) + G_{n1}(x_1) G_{n2}(x_1,x_1) 
\right. 
\nonumber \\
&& 
\left.
+ G_{n2}(x_1,1) G_{n2}(1,x_1) 
\right]
+ \pb G_{n1}(x_1) G_{n1}(x_1)
\nonumber \\
\hspace{-12mm} 
G_{n+1,2}(x_1,{x_2}) &=& \pb \left[  
G_{n1}(x_1) G_{n2}(x_1,{x_2}) + G_{n2}(x_1,{x_2}) G_{n1}({x_2}) \right.
\nonumber \\
&& \left. + G_{n2}(x_1,1) G_{n2}(1,{x_2}) 
\right]. 
\label{eq:farey_gen}
\ee
Here, $G_{n1}(\cdot)$ and $G_{n2}(\cdot,\cdot)$ in equation~\eqref{eq:farey_gen} 
represent an open and closed hyperedge, respectively. 
Each term corresponds to one realization of the hyperedges and the shortcut edges, 
and every realization appears only once. 
The unity in the argument is related to 
an old root that is not connected to any new roots.

From equation~\eqref{eq:farey_gen}, we obtain the recursion equations for 
\be
\vg_n 
= \left( \begin{array}{c} g_{n1} \\ g_{n2} \end{array} \right)
= \left( \begin{array}{c} G_{n1}(1) \\ G_{n2}(1,1) \end{array} \right)
\\ 
\begin{array}{rcl}
	\mrm{and} \quad \HH_n(x) &=& {}^t\left( H_{n1}(x),\cdots,H_{n4}(x) \right) 
	\\ &=& 
	{}^t\left( G_{n1}(x), G_{n2}(x,x), G_{n2}(x,1), G_{n2}(1,x) \right) 
\end{array}
\ee
as 
\be
\vg_{n+1} &=& \ff(\vg_n) = \left(
\begin{array}{c}
	p( g_{n1} + g_{n2} )^2 + \pb g_{n1}^2 \\ 
	\pb ( 2 g_{n1} g_{n2} + g_{n2}^2 ) 
\end{array}
\right), 
\label{eq:RG_Farey}
\\
\HH_{n+1} &=& \FF\left( \vg_n, \HH_n \right) 
= \left(
\begin{array}{c}
	H_{n1}^2 + p( 2 H_{n1} H_{n2} + H_{n3} H_{n4} ) \\ 
	\pb ( 2 H_{n1} H_{n2} + H_{n3} H_{n4} ) \\ 
	\pb H_{n3} ( H_{n1} + g_{n1} + g_{n2} ) \\
	\pb H_{n4} ( H_{n1} + g_{n1} + g_{n2} ) 
\end{array}
\right) .
\label{eq:F_farey}
\ee
Here, we omit the arguments ``$(x)$'' for all components of $\HH_n$.
Note that equation~\eqref{eq:RG_Farey} is obtained by substituting $x=1$ 
into equation~\eqref{eq:F_farey}. 
Equation \eqref{eq:F_farey} leads to 
\be
M_n = \left(
\begin{array}{cccc}
	2 H_{n1} + 2 p H_{n2} & 2 p H_{n1} & p H_{n4} & p H_{n3} \\
	2 \pb H_{n2} & 2 \pb H_{n1} & \pb H_{n4} & \pb H_{n3} \\
	\pb H_{n3} & 0 & \pb( H_{n1} + 1) & 0 \\
	\pb H_{n4} & 0 & 0 & \pb( H_{n1} + 1) \\
\end{array}
\right).
\label{eq:matrix_farey}
\ee
Again, we omit the argument ``(1)''.

By solving $\ff(\vg_n)=\vg_n$, the RG FPs are obtained as 
\be
g_{n1} = 1 - g_{n2} = \frac{1\pm |1-2p|}{2\pb} = 1, \, \frac{p}{\pb}. 
\ee
A transcritical bifurcation occurs at $p = 1/2$. 
For the initial condition: $g_{01} \in (0,1)$, 
$g_{\infty 1}$ equals $p/\pb$ for $p<\pctwo$ and 1 for $p \ge \pctwo$. 
By substituting this and 
\be
H_{\infty1}(1) = g_{\infty1}, \quad
H_{\infty2}(1) = H_{\infty3}(1) = H_{\infty4}(1) = 1 - g_{\infty1} 
\ee
into equation~\eqref{eq:matrix_farey}, 
the eigenvalues of $M_\infty$ are obtained as 
\be
\lambda_\infty = \left\{
\begin{array}{ccc}
	\lambda_+ , 1, 2p, \lambda_-  & \mrm{for} & p < 1/2
	\\
	2, 2\pb & \mrm{for} & p \ge 1/2
\end{array}
\right.
, 
\nonumber \\
\label{eq:Farey_lambda}
\mrm{where} \ \lambda_\pm := \dps \frac{1}{2} \left[ 1+4p \pm \sqrt{1+8p^2(1-2p)/\pb} \right].
\ee
Here $2\pb$ for $p \ge 1/2$ is triply degenerated. 
Because $\lambda_\infty^{(1)}=\lambda_+ \in [1,2)$ ($\iff \psi \in [0,1)$) for $p<1/2$ 
and $\lambda_\infty^{(1)}=2$ ($\iff \psi = 1$) for $p \ge 1/2$, 
we have $\pcone=0$ and $\pctwo=1/2$ for this PHG. 
For $p<\pctwo$, $\psi$ is expanded by $\pctwo-p$ as 
\be
\psi = \log_2( \lambda_+ ) = 1 - (8/\ln 2) (\pctwo - p)^2  + O\left( (\pctwo-p)^3 \right). 
\label{eq:psi_farey}
\ee


\section{Relation between the RG parameter and the fractal exponent}

\subsection{local fractal exponent}

By using the combining matrix, $\sigma_n := E(|C_{no}|)$ is given by  
\be
\sigma_n 
= \vec{c} \vec{D}_n, \quad 
\vec{D}_n := \left. \frac{d \HH_n}{dx} \right|_{x=1} 
= \left[ \prod_{m=0}^{n-1} M_m \right] \left. \frac{d \HH_0}{dx} \right|_{x=1}. 
\label{eq:sigma}
\ee
Here $\vec{c}$ is a certain constant row vector. 
Then, we have 
\be
\sigma_{n+1} 
= \vec{c} M_n \vec{D}_n
= \vec{c} \sum_{i}^{\Nsgl} M_n \vv_n^{(i)} \uu_n^{(i)} \vec{D}_n
= \vec{c} \sum_{i=1}^{\Nsgl} \lambda_n^{(i)} \vv_n^{(i)} \uu_n^{(i)} \vec{D}_n. 
\ee
Here we inserted an identity matrix $\sum_{i=1}^{\Nsgl} \vv_n^{(i)} \uu_n^{(i)}$. 
From the fact that $\vec{D}_n$ becomes parallel to $\vv_n^{(1)}$ for $n \to \infty$ 
and $\uu_n^{(i)} \vv_n^{(j)} = \delta_{ij}$, we have 
\be
\sigma_{n+1} \approx \lambda_n^{(1)} \sigma_n \quad \mrm{for} \quad n \gg 1.
\ee
Consequently, the order parameter is expressed as 
\be
m \propto \lim_{k \to \infty} \frac{1}{\Ncp^{k}} \prod_{n=0}^{k-1} \lambda_n^{(1)} 
= \exp \left[ - \ycp \sum_{n=0}^\infty (1 - \psi_n) \right], 
\nonumber \\
\psi_n := \log_\Ncp \lambda_n^{(1)}, 
\quad
\ycp := \ln \Ncp. 
\label{eq:psi_n}
\ee
We call $\psi_n$ a {\it local} fractal exponent, which satisfies $\lim_{n\to \infty}\psi_n=\psi$. 
Particularly, $\lim_{n \to \infty} 1-\psi_n = 0$ for $p > \pctwo$. 
Equation~\eqref{eq:psi_n} gives the relation between $\psi_n$ and $m$. 
Thus, if we know the relation between $\psi_n$ and the RG solution $\vg_n$, 
we can understand the relation between the RG behavior and the singularity of the order parameter.

\subsection{perturbation near the continuous bifurcation point}

In the following, we assume that all the roots of the PHG is almost surely open for $n \to \infty$ and $p>\pctwo$; 
\be
g_{\infty i} = \delta_{i1} 
(\iff H_{\infty j}(1) = \delta_{j1} )
\quad \mrm{for} \quad p > \pctwo. 
\label{eq:empirical1}
\ee 
When this holds, we say that the PHG has a {\it tight-root-connection (TRC)}.
For a backbone-connected PHGs, $\GGG_{n+1}$ is open if all $\GGG_n^{(\cdot)}$'s in it are open. 
Thus the FP $g_{\infty i} = \delta_{i1}$ exists for any $p$ although it may not be stable. 
The SBC PHG with $\Nrt=2$ always has a TRC.

In the case that $g_{n1}$ undergoes a continuous bifurcation at $p=\pctwo$, 
$1-\psi_n$ is expanded by $1-g_{n1}$, and the leading term is expressed as 
\be
1 - \psi_n = c_\mu (1-g_{n1})^\mu 
\quad \mrm{for} 
\quad |p - p_{c2}| \ll 1\quad \mrm{and} 
\quad 1 - g_{n1} \ll 1, 
\label{eq:mu}
\ee
where $c_\mu$ is a constant. 
By taking the limit of $n \to \infty$ for $p<\pctwo$, we have 
\be
1 - \psi = c_\mu (1 - g_{\infty 1})^\mu 
\quad \mrm{for} \quad 1 - g_{\infty 1} \ll 1. 
\label{eq:mu0}
\ee
Although $\mu$ is naturally equal to 1 in general, 
$\mu$ often equals 2 like Equation~\eqref{eq:psi_farey} 
in the previously known systems as shown in table~\ref{tab:graph}. 
We find the following 
\begin{itemize}
	\item If a PHG is SBC and has a TRC, $\mu > 1$.
\end{itemize}
We derive this in \ref{sec:absence} because it is a little long.

\section{Singularity of the order parameter}

\begin{figure}[t]
	\begin{center}
		\includegraphics[trim=40 320 70 40,scale=0.7,clip]{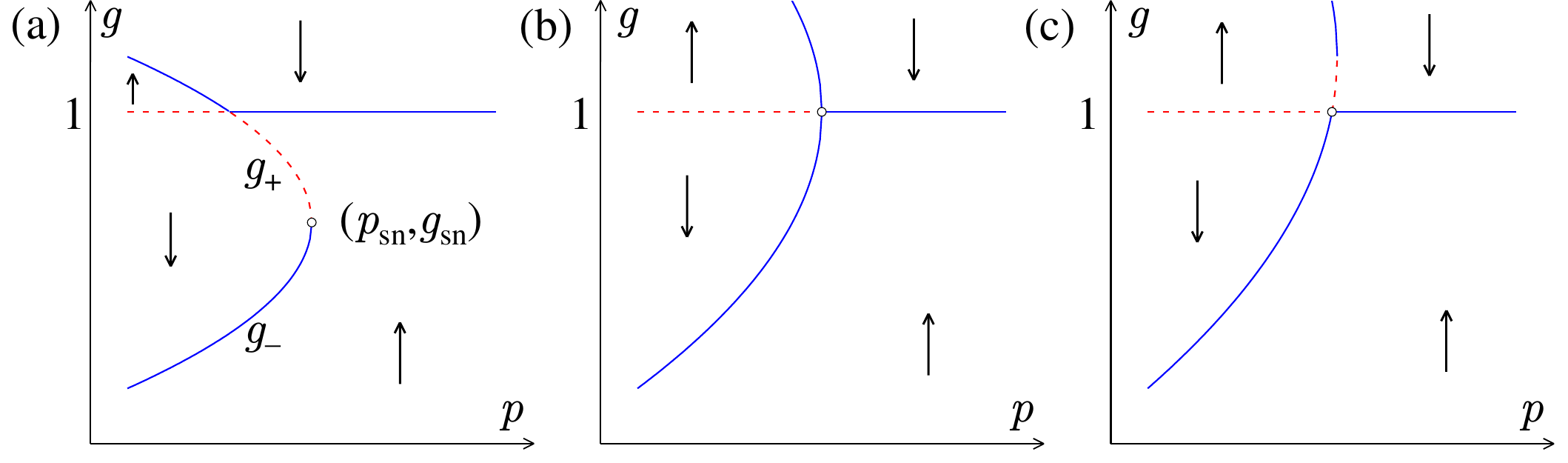}
	\end{center}
	\vspace{-5mm}
	\caption{\label{fig:bifurcation}
		The fixed-points of equation~\eqref{eq:RG-p} 
		as a function of the open-bond probability in the cases: 
		(a) $g_\sn \in (0,1)$, (b) $g_\sn=1$ and (c) $g_\sn>1$. 
		The solid and dashed lines denote the stable and the unstable fixed points, respectively. 
		The arrows denote the flow direction of the renormalization transformation of $g$. 
	}
\end{figure}

Let us consider the RG equation for $\vep:=p-\pctwo \ll 1$. 
When $|\vg_{n+1}-\vg_n| \ll 1$, the RG equation is given by $d\vg/dn = \ff(\vg) - \vg$. 
Hereafter, we consider the analytic continuation 
of $g_{n1}$ to a real-variable function $g(n)$. 
The basic types of bifurcation are covered by the differential equation 
\be
\frac{dg}{dn} = A (1 - g) [ (g - g_\sn)^2 + B (p - p_\sn) ],\ 
A>0,\ B>0, \ p_\sn \in (0,1), 
\label{eq:RG-p}
\ee
which yields the FPs 
\be
g(n) = 1, g_\pm, \quad 
g_\pm := g_\sn \pm \sqrt{B (p_\sn-p)}.
\ee
Here, $g_+$ and $g_-$ vanish for $p > p_\sn$. 
As shown in figure~\ref{fig:bifurcation}, the FP exhibits 
(i) a saddle-node bifurcation for $g_\sn \in (0,1)$, 
(ii) a pitchfork bifurcation for $g_\sn = 1$ 
and (iii) a transcritical bifurcation for $g_\sn > 1$. 
By using the solution of equation~\eqref{eq:RG-p}, we evaluate the order parameter as 
\be
\hspace{-6mm}
m \propto \exp \left\{ - \ycp \int_0^\infty dn [1-\psi(n)] \right\} 
= \exp \left\{ - \ycp \int_0^\infty dn c_\mu [ 1 - g(n) ]^\mu \right\}. 
\label{eq:m-psi_n}
\ee

\subsection{saddle-node bifurcation: $g_\sn \in (0, 1)$}

First, we consider the case that the saddle-node bifurcation point (SNBP): 
$(p_\sn, g_\sn)$ exists in the physical region. 
The transition point of $g(\infty)$ depends on the initial condition $g(0)$, 
which is a monotonically increasing function of $p$. 
If $g(0) > g_\sn$ at $p=p_\sn$, $g(0)$ and the unstable FP $g_+$ have a crossing point at $p<p_\sn$. 
This is the transition point, below which $g(\infty)=g_-$ and above which $g(\infty)=1$. 
If $g(0) \le g_\sn$ at $p=p_\sn$, on the other hand, 
the transition point is always $p=p_\sn$, 
below which $g(\infty)=g_-$ and above which $g(\infty)=1$. 
In both cases, $\psi$ exhibits discontinuous change as well as $g(\infty)$.

\subsubsection{[the case of $g(0) > g_\sn$ at $p=p_\sn$]}

First, we consider the case that the transition corresponds to the unstable FP. 
The transition point $\pctwo$ is given by solving $g_+=g(0)$ with respect to $p$.
Let $g_c$ be the value of $g_+$ at $p=\pctwo$. 
For $0 < \vep \ll 1$, $g(n)$ moves from the point slightly above the unstable FP: $g_c$ 
to the stable FP: 1. 
The RG equation~\eqref{eq:RG-p} is rewritten as 
\be
dz/dn = a (z + b \vep) + [O(z)+O(\vep)]^2, 
\label{eq:dif_sn-1}
\ee
where $z(n) := g(n)-g_c$, $a := 2 A(1-g_c)(g_c-g_\sn)$ and $b:=B/2(g_c-g_\sn)$. 
By integrating this, we have 
\be
a \left[ n - n^*(\vep) \right] = \int_{z^*}^{z(n)} \frac{dz}{z+b\vep}
= \left[ \ln \left(z + b \vep \right) \right]_{z^*}^{z(n)}
\nonumber \\
\iff z(n) = \left( z^* + b \vep \right) e^{a\left[n-n^*(\vep)\right] } - b \vep. 
\label{eq:sol_us}
\ee
Here, we introduce a constant $z^*$ such that $z(0)=O(\vep) \ll z^* \ll 1$ 
and $n^*(\vep)$ that satisfies $z(n^*(\vep)) = z^*$. 
By substituting $g(0) = g_c + \left. \frac{\del g(0)}{\del p} \right|_{p=p_c} \vep$, 
we have 
\be
n^*(\vep) = - \frac{1}{a} \ln \left[ 
\left( b + \left. \frac{\del g(0)}{\del p} \right|_{p=p_c} \right) 
\frac{\vep}{z^* + b \vep} \right]  
= -\frac{\ln \vep}{a} + O(\vep^0). 
\ee
Here $z(n)$ behaves like a sigmoid function that 
moves from 0 to $1-g_c$ around $n^*$ 
and the width of the crossover is $O(|n^*|^0)$. 
Because $n^*(\vep)$ diverges as $\vep \to 0$, we have 
\be
z(n) = z(n^*(\vep) \hat{n}) 
\xrightarrow[\vep \to 0]{} (1-g_c) \Theta( \hat{n}-1 )
\ee
where $\hat{n} := n/n^*(\vep)$ and $\Theta(\cdot)$ is a step function. 
Similarly, $\psi(n^*(\vep) \hat{n})$ converges to a step function 
jumping from $\psi_c$ to 1 at $\hat{n}=1$, 
where $\psi_c$ is the fractal exponent at the unstable FP $g(n)=g_c$. 
By using this, 
\be
\hspace{-14mm}
\int_0^\infty dn [ 1 - \psi(n) ] 
= n^*(\vep) \int_0^\infty d\hat{n} ( 1- \psi_c ) \Theta(1-\hat{n}) 
= -\frac{\ln \vep}{a} ( 1- \psi_c ) + O(\vep^0). 
\ee
From equation~\eqref{eq:m-psi_n}, 
the order parameter exhibits a power-law singularity as 
\be
m \propto \vep^{\beta_\sn}, \quad 
\beta_\sn := \frac{ \ycp  }{2A(g_c-g_\sn)} \frac{1-\psi_c}{1-g_c}. 
\ee
This transition is governed by an unstable FP, 
whose mechanism is essentially same with that of the ordinary second-order transition.

\subsubsection{[the case of $g(0) \le g_\sn$ at $p=p_\sn$]}

If $g(0) \le g_\sn$, $\pctwo$ equals $p_\sn$. 
In the vicinity of the SNBP, equation~\eqref{eq:RG-p} is approximated as 
\be
dz/dn = - a ( z^2 + \zeta^2) \quad \mrm{for} \quad |z| \ll 1
\label{eq:saddle-node-eq}
\ee
where $z(n) := g(n) - g_\sn$, $a:=(1-g_\sn)A$ and $\zeta^2 := B \vep$. 
For $0 < \vep \ll 1$, $z(n)$ starts from $z(0) \le 0$, spends long time around 0 
and converges to $1-g_\sn$. 
By integrating this, we obtain 
\be
- a( n-n^*) = \int_{z^*}^{z(n)} \frac{dz}{z^2+\zeta^2} 
= \frac{1}{\zeta} \left[ \arctan \frac{\zeta}{z} \right]_{z^*}^{z(n)}, 
\nonumber \\
\iff z(n) = \zeta \tan \left[ a (n - n^*) \zeta + \arctan(z^*/\zeta) \right].
\ee
Here we introduce $z^*$ and $n^*$ such that $0 < z^* \ll 1$ and $z(n^*)=z^*$. 
By substituting $n=0$ into this, we have
\be
n^* = \frac{1}{a\zeta} 
\left[ \arctan\frac{z^*}{\zeta} + \arctan\frac{|z(0)|}{\zeta} \right] 
= \frac{ \pi }{ a \zeta } + O(\zeta)  
\ee
for $z(0)<0$. [$n^*=\pi/2 a \zeta$ for $z(0)=0$. ]
Similarly to the argument for $g(0)>g_\sn$, 
$\psi(n^* \hat{n})$ converges to the step function jumping from $\psi_\sn$ to 1 at $\hat{n}=1$. 
Thus, we have 
\be
\int_0^\infty dn [ 1 - \psi(n) ] = \frac{\pi}{a\zeta} ( 1 - \psi_\sn ),  
\ee
where $\psi_\sn$ is the fractal exponent at the SNBP. 
(Here the contribution from the region where $z < -z^*$ is negligible.)
Finally, we obtain an essential singularity as 
\be
m \propto \exp[ - \alpha_\sn \vep^{-1/2} ], \quad 
\alpha_\sn := \frac{\pi \ycp}{A \sqrt{B}} \frac{1-\psi_\sn}{1-g_\sn} .  
\ee

\subsection{pitchfork bifurcation: $g_\sn = 1$}

For $g_\sn=1$, $\pctwo$ equals $p_\sn$ as far as $g(0) \in (0,1)$. 
The RG equation~\eqref{eq:RG-p} is rewritten as 
\be
dx/dn = - A x ( x^2 + \zeta^2 ), 
\ee
where $x := 1 - g$ and $\zeta^2 := B \vep$. 
This is integrated as 
\be
-A n = \int_{x(0)}^{x(n)} \frac{dx}{x(x^2+\zeta^2)}
= - \frac{1}{2 \zeta^2} 
\left[ \ln \left(1 + \frac{\zeta^2}{x^2} \right) \right]_{x(0)}^{x(n)} 
\nonumber \\ 
\begin{array}{rcl}
	\iff x(n) &=& \zeta \left[ \left( 1 + \frac{\zeta^2}{x(0)^2} \right) e^{2A \zeta^2 n} - 1 \right]^{-1/2} 
	\label{eq:x_n_marge}
	\\
	& \approx & \left\{ 
	\begin{array}{cl}
		(2An)^{-1/2} & \mrm{for} \ \  1/Ax(0)^2 \ll n \ll 1/A\zeta^2 \\
		\zeta e^{-A\zeta^2 n} & \mrm{for} \ \ n \gg 1/A \zeta^2
	\end{array}
	\right. .
\end{array}
\ee
Here we assume that $\zeta \ll x(0)$. 

\subsubsection{[the case of $\mu=1$]}

By using 
\be
\int_0^\infty \frac{e^{- \kappa n} dn}{\sqrt{1+\ell-e^{-2 \kappa n}}}
= \left[ \arcsin\left( \frac{\sin \theta}{\sqrt{1+\ell}} \right) \right]_0^{\pi/2}
= \frac{1}{\kappa} \left[ \frac{\pi}{2} + O\left( \sqrt{\ell} \right) \right], 
\ee
we have an essential singularity as 
\be
m \propto \exp \left[ - \ycp c_1 \zeta \frac{\pi}{2} \frac{1}{A \zeta^2} \right]
= \exp[-\alpha_\mrm{pf} \vep^{-1/2}], \quad
\alpha_\mrm{pf} := \frac{\pi}{2} \frac{\ycp c_1}{A \sqrt{B}}. 
\ee

\subsubsection{[the case of $\mu=2$]}

By using 
\be
\int_0^\infty \frac{e^{- \kappa n} dn}{1 + \ell - e^{- \kappa n} }
= \left[ \frac{\ln(1 + \ell - e^{-\kappa n})}{\kappa} \right]_0^\infty
= - \frac{1}{\kappa} \left[ \ln \ell + O(\ell) \right] , 
\label{eq:integration1}
\ee
we have a power-law singularity as 
\be
m \propto \exp\left[ \ycp c_2 \zeta^2 \frac{1}{2 A \zeta^2} \ln \frac{\zeta^2}{x(0)^2} \right]
\propto \vep^{\beta_\mrm{pf}}, 
\quad
\beta_\mrm{pf} := \frac{\ycp c_2}{2 A}.
\ee
Note that this singularity is governed by a stable FP 
and its origin is essentially different from the power-law singularity 
in the conventional second-order transition. 

\subsection{transcritical bifurcation: $g_\sn > 1$}

For $g_\sn > 1$, the transition point is given by 
$p_{c2} = p_\sn - (g_\sn-1)^2/B$ as far as $g(0) \in (0,1)$. 
In the vicinity of this point, the RG equation is expressed as 
\be
dx/dn = - a x( x + b \vep ) + O(x^3), 
\ee
where $x(n) := 1-g(n)$, $a := 2A(g_\sn-1)>0$ and $b:=AB/a$. 
This is integrated as 
\be
- a n = \int_{x(0)}^{x(n)} \frac{dn}{x(x + b \vep)} 
= - \frac{1}{b \vep} \left[ \ln \left( 1 + \frac{b \vep}{x} \right) \right]_{x(0)}^{x(n)} 
\nonumber \\ 
\begin{array}{rcl}
	\iff x(n) &=& b \vep \left[ \left( 1 + \frac{b \vep}{x(0)} \right) e^{ ab \vep n} - 1 \right]^{-1}
	\\
	& \approx & \left \{
	\begin{array}{ll}
		1/an & \mrm{for} \ \  1/ax(0) \ll n \ll 1/ab\vep \\
		b \vep e^{-ab \vep n} & \mrm{for} \ \ n \gg 1/ab\vep
	\end{array}
	\right. .
\end{array}
\label{eq:x_n_exp}
\ee


\subsubsection{[the case of $\mu=1$]}

By using equation~\eqref{eq:integration1}, we have a power-law singularity as 
\be
m \propto \exp \left[ \ycp c_1 b \vep \frac{1}{ab \vep} \ln \frac{b \vep}{x(0)} \right]
\propto \vep^{\beta_\mrm{tc}}, \quad 
\beta_\mrm{tc} := \frac{\ycp c_1}{2 A (g_\sn-1)}. 
\ee

\subsubsection{[the case of $\mu=2$]}

By using 
\be
\int_0^\infty \frac{ (e^{-\kappa n})^2 dn }{ [ 1 + \ell - e^{-\kappa n}]^2 }
&=& -\frac{1}{\kappa} \left[
\frac{e^{-\kappa n}}{1 + \ell - e^{-\kappa n}} + \ln( 1 + \ell - e^{-\kappa n})
\right]_0^\infty 
\nonumber \\
&=& \frac{1}{\kappa} \left[ \frac{1}{\ell} + \ln \ell + O(\ell) \right], 
\ee
we have 
\be
\int_0^\infty dn x(n)^2 \approx \left( b\vep \right)^2 
\frac{1}{ab \vep} \left[ \frac{x(0)}{b \vep} + \ln \frac{b \vep}{x(0)} \right]
= \frac{x(0)}{a} + \frac{b \vep}{a} \ln \frac{b \vep}{x(0)}. 
\ee
Consequently, we obtain an abrupt singularity as 
\be
m \propto \exp \left[ -\ycp c_2 \frac{b \vep }{a} \ln \frac{b \vep}{x(0)} \right]
= \left(\frac{\vep}{\vep_0}\right)^{- \gamma_\mrm{tc} \vep}, 
\quad \gamma_\mrm{tc} := \frac{\ycp c_2 B}{4A(g_\sn-1)^2}, 
\ee
where $\vep_0 := x(0)/b$.
(This is a decreasing function for $\vep > \vep_0/e$, which is actually an off-critical region.) 
In the limit of $\vep \searrow 0$, $m$ converges to a positive constant. 
But this is not like the ordinary first-order transition; 
$\mrm{d}m/ \mrm{d}p$ 
weakly diverges being proportional to $\ln(\vep_0/\vep)$. 


\section{Summary and discussion}

\renewcommand{\arraystretch}{1.5}
\begin{table}[t]
	\begin{center}
		\begin{tabular}{l||l|l|l}
			\hline 
			bifurcation & condition & singularity & exponent 
			\\
			\hline
			saddle-node & $g(0)>g_\sn$ 
			& power-law \ $\vep^{\beta_\sn}$ 
			& $\beta_\sn := \dps\frac{ \ycp  }{2A(g_c-g_\sn)} \dps\frac{1-\psi_c}{1-g_c}$
			\\ \cline{2-4} 
			($g_\sn \in (0,1)$) & $g(0)<g_\sn$ 
			& essential \ $e^{-\alpha_\sn \vep^{-1/2}}$ 
			& $\alpha_\sn := \dps\frac{\pi \ycp}{A\sqrt{B}} \dps\frac{1-\psi_\sn}{1-g_\sn}$
			\\ \hline \hline
			pitchfork & $\mu=1$ & essential \ $e^{-\alpha_\mrm{pf} \vep^{-1/2}}$ 
			& $\alpha_\mrm{pf} := \dps\frac{\pi}{2} \dps\frac{\ycp c_1}{A \sqrt{B}}$ 
			\\ \cline{2-4} 
			($g_\sn = 1$)  & $\mu=2$ & power-law \ $\vep^{\beta_\mrm{pf}}$ 
			& $\beta_\mrm{pf} := \dps\frac{\ycp c_2}{2 A}$ 
			\\ \hline \hline
			transcritical & $\mu=1$ & power-law \ $\vep^{\beta_\mrm{tc}}$ 
			& $\beta_\mrm{tc} := \dps\frac{\ycp c_1}{2 A (g_\sn-1)}$ 
			\\ \cline{2-4} 
			($g_\sn > 1$)  & $\mu=2$ & abrupt \ $\vep^{-\gamma_\mrm{tc} \vep}$ 
			& $\gamma_\mrm{tc} := \dps\frac{\ycp c_2 B}{4A(g_\sn-1)^2}$
			\\ \hline
		\end{tabular}
	\end{center}
	\caption{\label{tbl:summary}
		The correspondence 
		between the type of the bifurcation of the RG FP 
		and the singularity of the order parameter and critical exponents. 
	}
\end{table}
\renewcommand{\arraystretch}{1.0}


In this paper, we provided the generic theory 
of the bond percolation transition on PHGs to show the relation 
between the type of the singularity of the order parameter 
and the type of the bifurcation of the RG FP. 
The results are summarized in the table \ref{tbl:summary}.
In the case that the RG FP exhibits a saddle-node bifurcation, 
the singularity depends on the initial condition, 
which is given by the local connectivity in the minimum unit $\GGG_0$. 
In the case of the continuous bifurcations of the FP, 
the singularity depends on $\mu$: the order of the leading correction to the fractal exponent 
at the bifurcation point. 
All of the present results are checked to be correct for all PHGs in figure~\ref{fig:graph} 
by numerical calculation of equation~\eqref{eq:sigma} (not shown here).

We also showed that the sufficient condition for $\mu>1$ is 
that a PHG is SBC and has a TRC. 
The necessary condition is an open problem. 
As far as the PHGs in figure~\ref{fig:graph}, $\mu=2$ leads to SBC. 
Then, SBC may be also the necessary condition. 
On the other hand, TRC is not the necessary condition 
because the PHG (h) doesn't have a TRC, 
where both $P(R_n\sim [111])$ and $P(R_n\sim [112])$ are positive for $p>\pctwo$. 
In this case, however, no graph-operation is done on $r_{n3}$. 
Therefore, it can be eliminated from $R_n$, which yields the PHG (b). 
We don't know whether a nontrivial counter-example exists or not.





\appendix
\section{Sufficient condition for $\mu>1$}
\label{sec:absence}


Here we show that if a PHG is SBC and has a TRC, 
the first-order perturbation of the eigenvalue of the combining matrix at the FP equals zero at $p=\pctwo$ as 
\be
\delta \lambda_n^{(1)} 
:= \uu_{\infty}^{(1)} \delta M_{n} \vv_{\infty}^{(1)}
=0,\quad
\nonumber \\
\delta M_n := M_n - M_\infty 
= \sum_{k=1}^\Nsgl \left[ \lim_{n\to \infty} \frac{\del M_{n}}{\del H_{n k}} \right] \delta H_{nk} 
+ \sum_{l=1}^\Ncnc \left[ \lim_{n\to \infty} \frac{\del M_{n}}{\del g_{n l}} \right] \delta g_{nl}.  
\label{eq:perturbation0}
\ee
Hereafter, we omit the argument ``(1)'' of $\HH_n$, 
and we regard $\HH_n(1)$ and $\vg_n$ in $M_n$ are independent variables 
although each component of the former equals one of the component of the latter.



\begin{figure}[t]
	\begin{center}
		\includegraphics[trim=0 400 60 30,scale=0.6,clip]{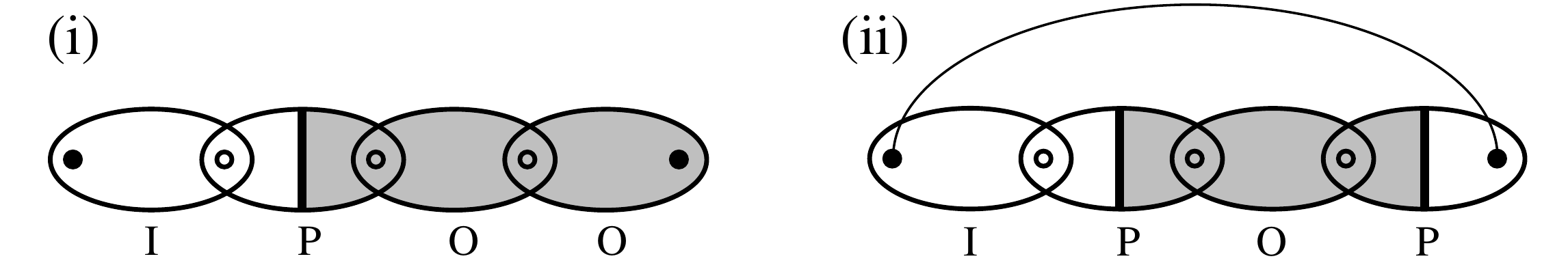}
	\end{center}
	\vspace{-5mm}
	\caption{\label{fig:perimeter}
		Examples of the inner structure of (i) a perimeter and (ii) an insider. 
		The characters: I, O and P denote insider, outsider and perimeter, respectively. 
		The shaded parts denote the components that are disconnected to $C_x$. 
	}
\end{figure}

As observed in Sec.~\ref{sec:farey1}, 
$G_{n+1,\omega}(x)$ is generally given by the summation of the terms 
that respectively represent the open-closed realizations of the hyperedges $\GGG_n^{(\cdot)}$'s 
and the lastly added shortcuts. 
Each $H_{ni}$ and $g_{ni}$ that appear in $H_{n+1,j}=F_j \left( \vg_n, \HH_n \right)$ 
also correspond to one of these hyperedges. 
These hyperedges in $H_{n+1,j}$ are classified into the following three categories 
in the relation with $C_x \subset V_{n+1}$: the cluster that corresponds to the argument of $\HH_{n+1}(x)$. 
\begin{itemize}
	\item {\it insider}\,: all roots are included in $C_x$, 
	e.g., $H_{n1}(x)=G_{n1}(x)$ and $H_{nj}(x)=G_{n\omega}(x,x,x)$. 
	Note that the roots are connected by $E_{n+1}$ 
	but they are not connected by $E_n^{(\cdot)}$ except $G_{n1}(x)$. 
	
	\item {\it outsider}\,: all roots are not included in $C_x$, 
	e.g., $g_{n1}=G_{n1}(1), \cdots, g_{n \Ncnc}=G_{n \Ncnc}(1,\cdots,1)$.
	
	\item {\it perimeter}\,: a part of the roots is included in $C_x$ and the others are not, 
	e.g., $H_{nj}(x)=G_{n\omega}(x,1,x)$.
\end{itemize}

If a PHG is SBC, every $\GGG_n^{(\cdot)} \subset \GGG_{n+1}$ includes 
at least one essential path that connects a pair of the roots. 
Therefore, the followings hold (see figure~\ref{fig:perimeter}). 
\begin{enumerate}
	\item A perimeter includes at least one perimeter of the previous generation. 
	
	\item If an insider includes the outsider of the previous generation, 
	there exist at least two perimeters of the previous generation.
\end{enumerate}

When the PHG has a TRC, 
every term in $M_{n,ij}=\del H_{n+1,i}/\del H_{nj}$ that {\it survives}, 
i.e., remains positive for $n \to \infty$, takes a form 
\be
p^a (\pb)^b \frac{\del}{\del H_{nj}} \left[ g_{n1}^k H_{n1}^{\Ncp-1-k} H_{nj} \right], 
\quad k \in \{0,1, \cdots, \Ncp-1 \}. 
\label{eq:diagram-term}
\ee
The prefactor $p^a (\pb)^b$ corresponds to whether each shortcut is open or not. 
Since at most one perimeter is allowed in the object of the differentiation, 
the property (ii) above leads to that 
\vspace{3mm} \\
\ (ii)$^\prime$ The surviving term in an insider includes no outsider of the previous generation; 
$k=0$ in equation~\eqref{eq:diagram-term}.


If $j=1$ in equation~\eqref{eq:diagram-term}, no perimeter exists. 
Furthermore $k$ equals zero from (ii)$^\prime$. 
Therefore, 
\be
M_{\infty,i1} 
= \Ncp \delta_{i1} \To v_{\infty j}^{(1)}=\delta_{j1}.
\label{eq:single_mode}
\ee
Here no $p$-dependent prefactor exists
because all open-closed realizations of the shortcuts contribute to it. 
In a similar argument, we obtain 
\be
\lim_{n\to \infty} \frac{\del M_{n,i1}}{\del g_{n l}} = 0 
\quad \mrm{for} \quad l \in \{1,\cdots,\Ncnc\}.
\label{eq:omake}
\ee


Let us divide the components of $\HH_n$ into two blocks: 
the insiders ($I$) and the perimeters ($P$). 
The dimensions are $\Ncnc$ and $\Nsgl-\Ncnc$, respectively. 
We can express the quantities defined in Sec.~\ref{sec:gen_fnc} and \ref{sec:comb} as 
\be
\hspace{-4mm}
\HH_n = \left( \begin{array}{c}
	\HHH_{nI} \\ \HHH_{nP} 
\end{array} \right), \quad
M_{n} = \left( \begin{array}{cc}
	\MMM_{n,II} & \MMM_{n,IP} \\
	\MMM_{n,PI} & \MMM_{n,PP}
\end{array} \right)
, \quad
\uu_n^{(1)} = (\UUU_{nI}, \UUU_{nP}). 
\ee
The properties (i) and (ii) lead to 
$\MMM_{\infty,PI}=0$ and $\MMM_{\infty,IP} = 0$, respectively. 
Consequently, $M_\infty$ is block-diagonal. 
Then, the largest eigenvalue of $\MMM_{\infty,II}$ equals that of $M_\infty$, namely, $\lambda_\infty^{(1)}=\Ncp$ 
and $\UUU_{\infty P}=\vec{0}$. 
Furthermore, the property (ii)$^\prime$ leads to that all the surviving terms in $\MMM_{\infty,II}$ 
take a form $p^a \pb^b H_{n1}^{\Ncp-1}$. 
Thus, we have 
\be
\lim_{n\to \infty} \frac{\del \MMM_{n,II}}{\del H_{n1}} = (\Ncp-1) \MMM_{\infty,II}
\nonumber \\
\quad \To \quad 
\uu_\infty^{(1)} \left[ \lim_{n\to \infty} \frac{\del M_n }{\del H_{n 1}} \right]
= \Ncp (\Ncp-1) \uu_\infty^{(1)} . 
\label{eq:sufcond}
\ee 
Here we used $\UUU_{\infty P}=\vec{0}$.


Small deviation from the FP at $p=\pctwo$: 
$\delta \vg_n := \vg_n - \vg_\infty$ and $\delta \HH_n := \HH_n - \HH_\infty$ 
obey to the recursion equation: 
\be
\delta \HH_{n+1} &=& \HH_{n+1} - \HH_\infty = \FF(\vg_n, \HH_n) - \FF(\vg_\infty, \HH_\infty) 
\nonumber \\
&=& M_\infty \delta \HH_n 
+ \sum_{l=1}^{\Ncnc} \left[ \lim_{m\to \infty} \frac{\del \FF}{\del g_{m l}} \right] \delta g_{nl} 
+ [O(|\delta \HH_n|)+O(|\delta \vg_n|)]^2. 
\ee
By multiplying $\uu_{\infty}^{(1)}$ from the left, we have 
\be
\uu_{\infty}^{(1)} \delta \HH_{n+1} = 
\Ncp [ \uu_{\infty}^{(1)} \delta \HH_{n} ] + \sum_{l=1}^{\Ncnc} 
\left[ \lim_{m \to \infty} \frac{ \del \uu_{\infty}^{(1)} \FF  }{\del g_{m l} } \right] \delta g_{n l} . 
\ee
The second term equals zero because $\UUU_{\infty P}=\vec{0}$ 
and, therefore, $\uu_{\infty}^{(1)} \FF$ includes no outsider.
Then, it holds that 
\be
\lim_{n \to \infty} \uu_\infty^{(1)} \delta \HH_n = 0. 
\label{eq:orthogonal_cond}
\ee
Otherwise, $|\delta \HH_n|$ would not converge to zero  for $n\to \infty$ 
but diverge as $\Ncp^n$. 

By using the results above, we obtain $\lambda_\infty^{(1)}=0$ as 
\be
\delta \lambda_n^{(1)} 
&=& \sum_{i,j=1}^\Nsgl u_{\infty i}^{(1)} \left\{ 
\sum_{k=1}^\Nsgl \left[ \lim_{n\to \infty} \frac{\del M_{n,ij}}{\del H_{n k}} \right] \delta H_{nk} 
+ 
\sum_{l=1}^\Ncnc \left[ \lim_{n\to \infty} \frac{\del M_{n,ij}}{\del g_{n l}} \right] \delta g_{nl} 
\right\}  v_{\infty j}^{(1)}
\nonumber \\
\nonumber \\
&=& \sum_{k=1}^\Nsgl \delta H_{nk} \sum_{i=1}^\Nsgl u_{\infty i}^{(1)} 
\left[ \lim_{n\to \infty} \frac{\del M_{n,ik}}{\del H_{n 1}} \right] 
\nonumber \\
&=& \Ncp (\Ncp-1) \sum_{k=1}^\Nsgl \delta H_{nk} u_{\infty k}^{(1)} 
\xrightarrow[n \to \infty]{} 0. 
\label{eq:perturbation}
\ee
Here we use equation~\eqref{eq:single_mode}, \eqref{eq:omake}, 
$ 
\frac{\del M_{ij}}{\del H_k} = \frac{\del^2 F_i}{\del H_j \del H_k} = \frac{\del M_{ik}}{\del H_j}, 
$ 
\eqref{eq:sufcond} and \eqref{eq:orthogonal_cond} in order.


Finally, let us check the above results in the case of the Farey graph. 
At $p=\pctwo$, we have 
\be
M_\infty = \left(
\begin{array}{cc|cc}
	2 & 1 & 0 & 0 \\
	0 & 1 & 0 & 0 \\ \hline
	0 & 0 & 1 & 0 \\
	0 & 0 & 0 & 1 
\end{array}
\right)
, \quad
\lim_{n \to \infty} \frac{\del M_n}{\del H_{n1}(1)} = \left(
\begin{array}{cc|cc}
	2 & 1 & 0 & 0 \\
	0 & 1 & 0 & 0 \\ \hline
	0 & 0 & 1/2 & 0 \\
	0 & 0 & 0 & 1/2
\end{array}
\right), 
\\
\lambda_\infty^{(1)}=2, \quad
\uu_\infty^{(1)} = (1,1,0,0), \quad 
\vv_\infty^{(1)} = {}^t(1,0,0,0), 
\\
\delta \HH_n(1) = {}^t( g_{n1}-1, g_{n2}, g_{n2}, g_{n2} ) 
= g_{n2} \times {}^t(-1,1,1,1). 
\ee
We can confirm that equations~\eqref{eq:single_mode}, \eqref{eq:sufcond} 
and \eqref{eq:orthogonal_cond} hold.

\section*{References}

\providecommand{\newblock}{}

\end{document}